\documentstyle[graphicx]{mn}

\def\etal{{et al.\ }}

\def\x2{$\chi^{2}$}

\def\asca{{\it ASCA }}

\def\x2{$\chi^{2}$}
\def\lunits{$\rm erg~s^{-1}$}
\def\funits{$\rm erg~cm^{-2}~s^{-1}$}
\def\cunits{$\rm cm^{-2}$}

\newbox\grsign \setbox\grsign=\hbox{$>$} \newdimen\grdimen \grdimen=\ht\grsign
\newbox\simlessbox \newbox\simgreatbox \newbox\simpropbox
\setbox\simgreatbox=\hbox{\raise.5ex\hbox{$>$}\llap
     {\lower.5ex\hbox{$\sim$}}}\ht1=\grdimen\dp1=0pt
\setbox\simlessbox=\hbox{\raise.5ex\hbox{$<$}\llap
     {\lower.5ex\hbox{$\sim$}}}\ht2=\grdimen\dp2=0pt
\setbox\simpropbox=\hbox{\raise.5ex\hbox{$\propto$}\llap
     {\lower.5ex\hbox{$\sim$}}}\ht2=\grdimen\dp2=0pt

\def\ASCA{{\it ASCA~}}
\def\ROSAT{{\it ROSAT~}}

\begin{document}

\title[ASCA observations of an obscured high-redshift QSO]
{ASCA observations of deep ROSAT fields IV. 
 IR and hard X-ray observations of an obscured high-redshift QSO
}

\author[I. Georgantopoulos et al.]
{\Large I. Georgantopoulos$^1$  O. Almaini$^2$ T. Shanks$^3$ G.C. Stewart$^4$ 
R.E. Griffiths$^5$  B.J. Boyle$^6$, K.F. Gunn$^3$  \\
$^1$ National Observatory of Athens, I. Metaxa \& B. Pavlou, Palaia Penteli, 
15236, Athens, Greece \\
$^2$ Institute for Astronomy, University of Edinburgh, Blackford Hill, 
Edinburgh, EH9 3HJ \\
$^3$ Physics Department, University of Durham, South Road, Durham, DH1 3LE \\
$^4$ Department of Physics and Astronomy, The University of Leicester, 
Leicester, LE1 7RH \\
$^5$ Department of Physics, Carnegie Mellon University, Wean Hall, 
5000 Forbes Ave., Pittsburgh, PA 15213, USA \\
$^6$ Anglo-Australian Observatory, PO Box 296, Epping NSW 2121, Australia \\
}

\maketitle
\begin{abstract}
We use UKIRT and \ASCA observations to determine the nature of  
a high redshift ($z=2.35$) narrow-line AGN, previously discovered 
by Almaini et al. (1995). The UKIRT observations 
show a broad $H_\alpha$ line  while no $H_\beta$ 
line is detected. This together with the red colour ($B-K=5.4$) 
 suggest that our 
object is a moderately obscured QSO ($A_V >3$), in   optical
wavelengths. 
The \ASCA data suggest a hard spectrum, 
probably due to a large obscuring column, 
with $\rm \Gamma=1.93^{+0.62}_{-0.46},  N_H\sim 10^{23} cm^{-2}$. 
The combined \ASCA and \ROSAT data again suggest a 
heavily obscured spectrum ($N_H\sim 10^{23}$ \cunits or $A_V\sim 100$).
In this picture, the
\ROSAT soft X-ray emission may arise 
from electron scattering, in a similar fashion to  
local Seyfert 1.9.  
Then, there is  a large discrepancy between 
the moderate reddening  witnessed by the IR and the 
large X-ray absorbing column. 
This could be possibly explained on the basis of 
 e.g. high gas metallicities or 
  assuming that the X-ray absorbing column 
 is inside the dust sublimation radius. 
 An alternative explanation can be obtained when we 
allow  for variability between the \ROSAT and 
\ASCA observations. Then the best fit spectrum is still flat, 
$\Gamma=1.35^{+0.16}_{-0.14}$, but with
low intrinsic absorption in better agreement with the IR data, while   
the \ROSAT normalization is  a factor of two below 
the \ASCA normalization.  
This object may be one of the bright examples of a type-II  
QSO population at high redshift, previously 
undetected in optical surveys. The hard X-ray spectrum 
of this object suggests that such a population could make 
a substantial contribution to the X-ray background.  

\end{abstract}

\newpage

\section{INTRODUCTION}
In recent  years, deep field observations with the X-ray missions 
\ROSAT and \ASCA have led to great progress in understanding the 
extragalactic X-ray source populations. 
\ROSAT surveys have resolved 70 per cent of the extragalactic light at soft 
X-ray energies reaching a surface density of 1000 $\rm deg^{-2}$
(Hasinger et al. 1998).
Spectroscopic follow-up observations showed that the 
majority of sources are broad-line QSOs at high redshift, $\rm z\sim1.5$,
(e.g. Shanks et al. 1991; Georgantopoulos et al. 1996; Schmidt et al. 1998). 
However, a large fraction of the X-ray sources at faint fluxes are 
associated with faint optical galaxies. These have narrow emission 
lines and X-ray luminosities ($L_x\sim 10^{42}$ \lunits) 
orders of magnitude above those of normal galaxies 
(Roche et al. 1995; Griffiths et al. 1996; McHardy et al. 1998).  
Their hardness ratios suggest a flat spectrum, $\Gamma \sim 1.5$,
(Romero-Colmenero et al. 1996; Almaini et al. 1996). 
There is strong evidence that these galaxies host an AGN as they 
present either high ionization lines or broad lines in their 
spectra (Schmidt et al. 1998). 

\ASCA provided us with the first images of the hard X-ray sky (2-10 keV). 
Due to its moderate spatial resolution it reaches a flux limit 
of $5\times 10^{-14}$ \funits, an order of magnitude above \ROSAT,    
resolving about 30 per cent of the 2-10 keV X-ray background (XRB),
(Georgantopoulos et al. 1997; Ueda et al. 1998). 
The number count distribution, $\log N-\log S$, is a factor of two 
above the \ROSAT counts, where the latter were converted to the 
2-10 keV band, using an average spectral index of the \ROSAT sources 
i.e. $\Gamma =2$. This immediately suggests the presence of a population  
with a flat X-ray spectrum other than the broad-line QSOs which   
have steep X-ray spectra in the \ROSAT band.   
Indeed, despite the difficulties in the optical follow-up 
due to the large position error box, there are now several examples 
of narrow-line AGN with hard X-ray spectrum at redshifts 
$\rm z<1$, (Ohta et al. 1996; Iwasawa et al. 1997; 
Akiyama et al. 1998; Boyle et al. 1998)  
These narrow line AGN may be identical to those detected 
in the \ROSAT surveys. Their flat X-ray spectrum would then 
make them easily detected at hard X-ray energies. 
They may represent the brightest examples of high 
luminosity, type-II QSOs, that is the 
high redshift counterparts of the Seyfert 1.9 and Seyfert 2 
galaxies. The existence of such a population  
has been debated in the last few years (e.g. Halpern et al. 1998). 
This population 
is likely to be significantly under-represented
in optical QSO surveys, 
 which are based on the presence of strong, usually blue,
 continua and broad emission lines.
Moreover, the flat X-ray spectra of \ASCA narrow-line AGN 
bear the promise that these belong to the long sought 
population which forms a substantial fraction of the 
XRB (e.g. Comastri et al. 1995; Madau et al. 1995).  

One of the most well-known examples of the narrow-line 
AGN population is RXJ13334+001 at a redshift of $z=2.35$.
This was discovered in our deep \ROSAT survey and its optical 
and \ROSAT X-ray spectrum were presented in Almaini et al. (1995).
It presents narrow $\rm Ly_\alpha$, C{\sc iv} emission lines on top of 
a strong UV continuum. Its \ROSAT X-ray spectrum appears to be hard, 
$\Gamma \sim 1.6$, with the column density  fixed to 
the Galactic value $N_H\approx 2.6\times10^{20}$ $\rm cm^{-2}$; 
the 95 per cent upper limit   
on its absorbing column is $N_H\sim 3 \times 10^{21}$
$\rm cm^{-2}$ when $\Gamma=2$.    
However, the  \ROSAT passband and  limited spectral 
resolution  did not allow a detailed study of its 
X-ray spectrum. Here, we present the \ASCA observations of this 
object extending the spectral coverage to 10 keV at the observer's frame. 
In addition,  we discuss observations at UKIRT, aimed at detecting 
broad emission lines (see also Shanks et al. 1995, 1996).

\section{The Infrared Observations} 

\begin{figure}
{\includegraphics[height=7.5cm]{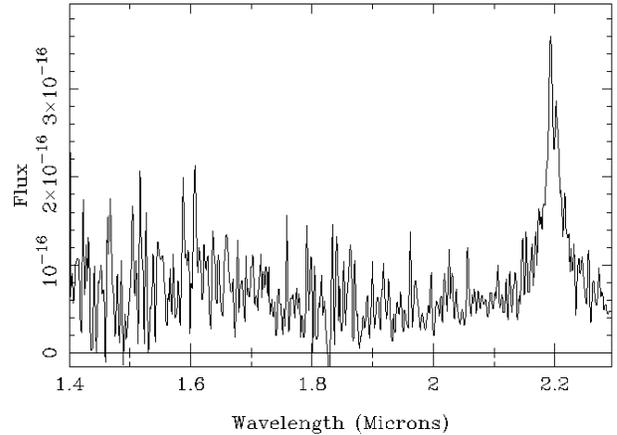}}
\caption{Near infrared spectrum of RXJ13334+0001 taken 
with the UK Infrared telescope in Hawaii.}
\label{ir2}
\end{figure}

During March and May 1995, observations of RXJ13334+0001 were made at
the UK Infrared Telescope in Hawaii. Photometry with IRCAM3 revealed a
point-like object in $0.6$ arcsec seeing with a magnitude
$K'=16.6$. This gives $B-K=5.4$ compared to $B-K\sim2.5$ for normal
QSOs at this redshift (Webster et al. 1995; Barvainis 1993), suggesting
$\sim3$ magnitudes of UV extinction (although, as discussed in Section
5, the internal extinction could be significantly larger if the UV
flux is dominated by surrounding star-forming activity rather than the
AGN).  An infrared spectrum was obtained with the aim of detecting
redshifted $H_\alpha$. A short $2000{\rm s}$ exposure was taken with
the CGS4 spectrograph using a 2.4 arcsec slit. The resulting spectrum,
with a resolution of $\sim25{\rm\AA}$ is shown in Fig. \ref{ir2}. The
broad emission line at 2.2 microns corresponds exactly to the expected
position of $H_\alpha$, confirming the redshift $z=2.35$.  The full
width at half maximum corresponds to $\sim 260{\rm \AA}$, i.e.
$\sim3500$km s$^{-1}$, which is entirely consistent with ordinary
broad-line QSOs. Hence the UKIRT observations confirm that
RXJ13334+0001 is a QSO obscured by a moderate amount of dust,
sufficient to obliterate the UV broad emission lines but not the broad
$H_\alpha$.

\begin{figure*}
{\includegraphics[height=8.0cm]{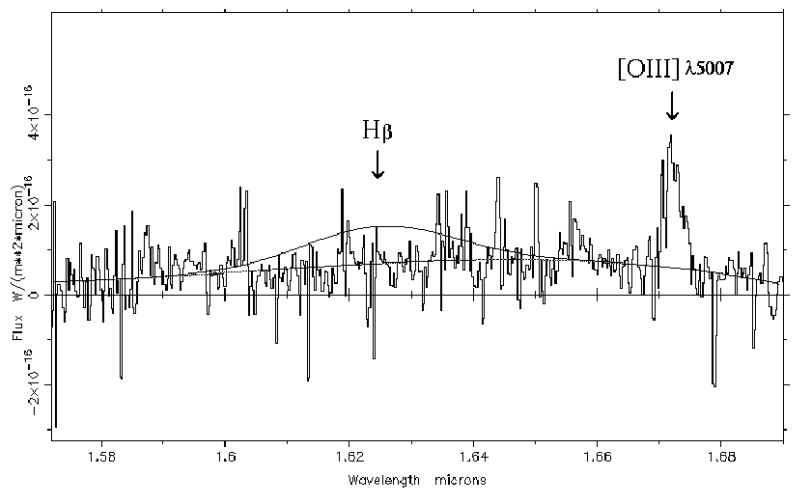}}
\caption{Additional near infrared spectrum of RXJ13334+0001, focussing
on the region near $H_\beta$. For comparison, the expected broad
line flux is shown above the continuum on the basis of a line ratio
$H_\alpha/H_\beta=3$.}
\label{ir1}
\end{figure*}

The short exposure displayed in Fig. \ref{ir2} was insufficient to
detect $H_\beta$ at 1.63 microns. Detection of this line would allow a
direct determination of the reddening from the Balmer decrement.  A
deeper service observation was therefore obtained during April 1996
using CGS4 with a 2.5 arcsec slit and a 150 lines/mm grating at 2nd
order, giving improved spectral resolution ($\sim 3{\rm\AA}$) over a
narrow wavelength range.  Broad $H_\beta$ was still not detected, but
this new spectrum allowed us to place a useful upper limit on the line
flux. This limit was obtained by fitting a model consisting of the
continuum plus a broad line with the same width as $H_\alpha$.  The
continuum level was well determined by fitting a 4th order polynomial
to the spectrum after excluding the [OIII] emission lines and a region
$100{\rm\AA}$ either side of the expected peak of $H_\beta$. By
varying the line flux and performing a $\chi^2$ fit to the data, we
obtain a limit $H_\alpha/H_\beta>8.4$ at the $99\%$ confidence level.
On the basis of Case B recombination (Baker \& Menzel 1938) one would
expect a line ratio $H_\alpha/H_\beta\simeq3$, which is clearly
inconsistent with the data (see Fig. \ref{ir1}).  Observed broad-line
QSOs have somewhat higher Balmer decrements than Case B predictions
(typically $H_\alpha/H_\beta\sim4$) but these are also inconsistent
with RXJ13334+0001.  We conclude from this limit that the broad line
region is undergoing photoelectric extinction, with a lower limit
$A_{\rm V}>3$.

The detection of broad $H_\alpha$ but non-detection of $H_\beta$ 
 formally classifies this
object as a QSO equivalent to `Type 1.9' Seyferts (Osterbrock 1981). The
service observation also revealed [OIII]$\lambda5007$ at 
$1.673$ microns (consistent with $z=2.341\pm0.005$) with an equivalent
width of $32{\rm \AA}$ and a FWHM of $540$km s$^{-1}$. Assuming a similar
width for narrow $H_\beta$ one can obtain a limit on the line ratio
[OIII]$/H_\beta>5.6$ (using the same technique described above). We note
that this is consistent with local Seyfert 2 galaxies (Veilleux \&
Osterbrock 1987). 

\section{THE X-RAY OBSERVATIONS}
This object was detected in the BJS864 field of our deep 
\ROSAT PSPC survey (Shanks et al. 1991; Georgantopoulos et al. 1996).
  There are two PSPC exposures available 
of 50 ksec total, the first obtained in January 1993 and the second in 
January 1994. 
The \ROSAT data were described in detail by Almaini et al. (1995). 
The 0.1-2 keV flux is  $4.5\times 10^{-14}$ \funits, translating to a 
luminosity of $L_x \sim 10^{45}$ \lunits ($\rm H_0=65 km~s^{-1}~Mpc^{-1}, 
q_0=0.5$) at the redshift of this object.   
Its optical position is 13h43m29.2s +00$^\circ$01{'}33{"} (J2000). 

RXJ13334+001 was observed by \ASCA (Tanaka, Inoue \& Holt 1994) 
in January 1996. 
The net exposure after screening the data using the 
standard criteria (REV1), (Yaqoob 1997), is 77 ksec
for the Gas Imaging Spectrometers (GIS; Tashiro et al. 1996) 
and  74 ksec for 
the Solid State Imaging Spectrometers (SIS; Gendreau 1995). 
The source was clearly detected by \ASCA with an 
 X-ray flux in the 2-10 keV band of $2.4\times 10^{-13}$ \funits.  
This translates to a luminosity of 
$L_x\sim 3\times 10^{45}$ \lunits in the same band 
($\rm H_0=65, q_0=0.5$). 
The SIS and GIS spectra were extracted from a region within 1.5 and 2 arcmin 
respectively of the X-ray centroid. Background data were 
accumulated from several source-free regions within the same field. 
The photons are grouped so that each channel contains 
a minimum of 20 counts (source plus background). 
 The spectral analysis was performed using the 
{\small XSPEC v10} package.

\section{The X-ray spectral results} 
\subsection{The ASCA fits}
The spectral fits to the \ROSAT data alone were presented in 
Almaini et al. (1995). Here, we present the spectral fits to the 
\ASCA GIS and SIS data. 
Following our \ROSAT results, we use a single power-law model with 
absorption in the  QSO rest-frame. An additional absorption 
component was set constant to the Galactic value ($\rm N_H=2.6\times 
10^{20} cm^{-2}$). 
The results are given in table 1: in column (1) we give the column density 
in units of $10^{22}$ $\rm cm^{-2}$; in column (2)  we give the photon 
index $\Gamma$; column (3) contains the value of the $\chi^2$ 
 divided by the number of degrees of freedom; finally, column (4) 
 contains the probability at which the fit is acceptable. 
The values presented with no associated error bars 
were kept constant during the spectral fit. 
All uncertainties given correspond to the 90 per cent 
confidence level. 

\begin{table}
\caption{Power-law fits to the \ASCA data.}
\begin{center}
\begin{tabular}{cccc}

 $N_H$ ($\rm 10^{22} cm^{-2}$) & $\Gamma$  & $\chi^2/d.o.f.$  & prob.  \\ \hline 
$20_{-16}^{+25}$& $1.93_{-0.46}^{+0.62}$  &34.3/39&0.69 \\
$10_{-7}^{+9}$& 1.56& 36.1/40 & 0.64 \\
0 & $1.37_{-0.27}^{+0.17}$&  39.6/40 & 0.49 \\
0 & 1.56  & 42.7/41 & 0.40 \\  

\end{tabular}
\end{center}
\end{table}
 
\begin{figure}
\rotatebox{270}{\includegraphics[height=8.5cm]{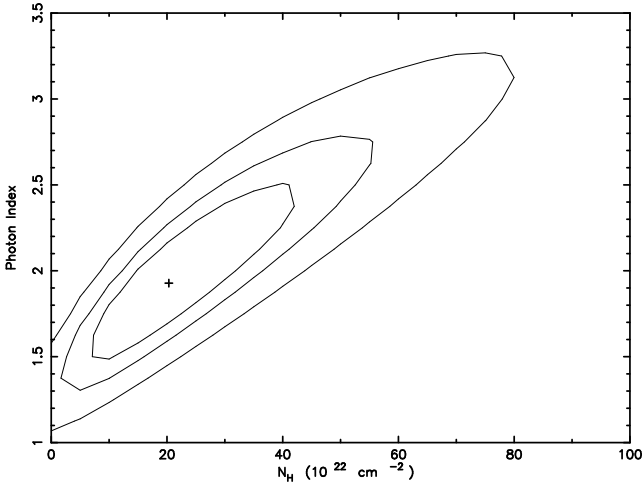}}
\caption{The confidence contours for the photon index and the 
absorbing column density for the \ASCA data alone. 
The contours correspond to the 
68, 90 and 99 per cent confidence levels.}
\label{cc1}
\end{figure}

The best-fit model has a steep photon index, $\Gamma=1.93^{+0.62}_{-0.46}$, 
albeit with a large uncertainty; this is 
consistent with the spectral slope of nearby AGN (Nandra \& Pounds 1994).  
However, the absorbing column is  large with $N_H\sim 20\times 10^{22}$ 
\cunits.
The confidence contours for these parameters are given in Fig. \ref{cc1}. 
It is evident that the \ASCA data favour a hard spectrum:
the error contours show that $\rm N_H$ and $\Gamma$ 
are correlated in the sense that 
we obtain either a steep power-law with a large absorbing column or 
a flat power-law with a moderate column. Thus, 
when we fix the column density to  $\rm N_H=0$, we obtain 
a flat spectral index $\Gamma=1.37^{+0.17}_{-0.27}$. 
Finally, we fix both the column density and the 
power-law to the values obtained from the 
\ROSAT data alone ($\Gamma=1.56, N_H=0$). 
We see that the \ROSAT best-fits are not rejected   
by the \ASCA data. 
In conclusion,  the \ASCA data favour a hard spectrum,
 in agreement with the \ROSAT data. 
However, with the current photon statistics it is not clear whether 
this hard spectrum is due to a flat spectral index or 
 a high amount of photoelectric absorption. 

\subsection{The ROSAT-ASCA joint fits}
Next, we combine the \ROSAT with the \ASCA data. Thus, we 
 extend the spectral coverage from 0.3-33 keV at the 
emitter's rest-frame and also we improve substantially the 
photon statistics. We first use a simple model, i.e.  a power-law model with 
absorption both at the emitter's rest-frame and the Galaxy. The 
Galactic column density is fixed at $\rm N_H=2.6\times 10^{20} cm^{-2}$ 
 as before. 
The results are presented in table 2: column (1) gives the 
column density at the emitter's rest frame in units 
of $10^{22}$ $\rm cm^{-2}$; column (2)  
gives the photon index;
 column (3) gives the covering factor in the case 
of the scattered power-law model (see below);
column (4) lists the $\chi^2$ and the number of degrees of freedom; while 
column (5) gives the probability that the model is acceptable. 

\begin{table*}
\caption{The spectral fits results to the joint ROSAT-ASCA data.}
\begin{center}
\begin{tabular}{lccccc}

 & $N_H$ {\small ($10^{22} \rm cm^{-2}$}) & $\Gamma$ & cov. factor
& $\chi^2/d.o.f.$  & prob.  \\ \hline 
{\bf Single power-law} ({\small ASCA-ROSAT norm. tied })
&$8_{-5}^{+9}$&$1.47_{-0.20}^{+0.43}$& -  &71.7/55&0.06 \\
{\bf Single power-law} ({\small ASCA-ROSAT norm. tied})& 0 & 
$0.96^{+0.14}_{-0.12}$ & - &  73.0/56& 0.06 \\
{\bf Single power-law} ({\small ASCA-ROSAT norm. untied}) &0  & $1.35_{-0.14}^{+0.16}$ & - & 59.4/55 & 0.32 \\
{\bf Scattered power-law} &
$35_{-13}^{+42}$&$2.10_{-0.17}^{+0.15}$& $0.90^{+0.05}_{-0.09}$ 
&50.4/53&0.67 \\ 

\end{tabular}
\end{center}
\end{table*}

\begin{figure}
\rotatebox{270}{\includegraphics[height=8.5cm]{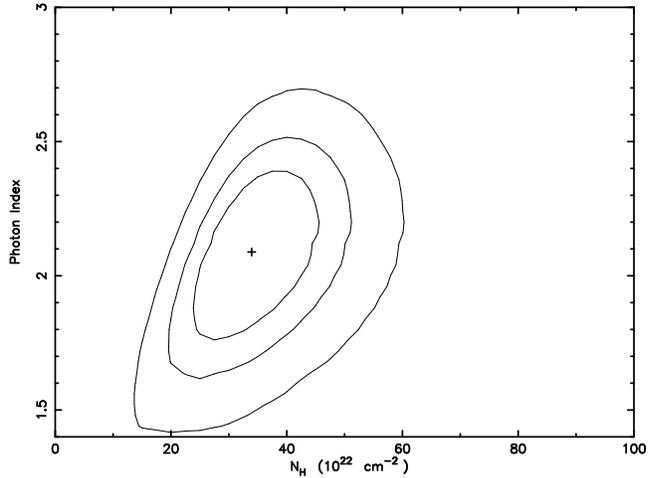}}
\caption{The confidence contours of the photon index and the column density
for the scattered power-law model, in the joint ROSAT-ASCA spectral fits. 
The contours correspond to the 68, 90 and 
99 per cent level of significance. 
}
\label{cc2}
\end{figure}

\begin{figure}
\rotatebox{270}{\includegraphics[height=8.5cm]{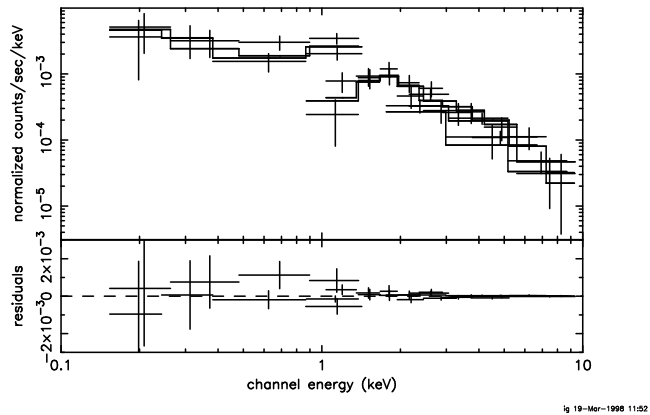}}
\caption{The ROSAT-ASCA spectrum. The solid lines  
represent the best-fit   model of a combination of an absorbed 
hard power-law and a scattered power-law at soft X-rays. 
}
\label{xspec}
\end{figure}

The best-fit photon index is hard  with an appreciable error,
while a large column density is again favoured by the data. 
Although, the best-fit model does not provide an excellent fit to the data,
it cannot be strongly rejected at a high level of significance 
(only at the $\sim 2\sigma$ confidence level). 
When we fix the column density to $N_H=0$ we obtain a very 
flat spectral index, $\Gamma \sim 1$. 
Next, we leave the \ASCA and \ROSAT normalizations untied. 
The reduced $\chi^2$ is significantly improved compared to the 
previous model (at over the 99.9 confidence level  
 for one additional parameter, according to the F-test, 
 Bevington \& Robinson 1995). 
However, the \ASCA normalization is higher than the \ROSAT normalization 
by a factor of two (3.4$\pm1.0$ and 1.7$\pm0.3$ 
$\times 10^{-5}$ $\rm photons~keV^{-1}~cm^{-2}~s^{-1}$ at 1 keV
respectively). 
  This difference is significant at the 99.9 confidence level.  
 It is highly unlikely that this discrepancy is due to 
calibration uncertainties in  the \ASCA or \ROSAT instruments. 
In contrast, it is likely that this could be due to flux variability
of our object  
between the \ASCA and the \ROSAT epochs. 
Note that the  underlying assumption in our spectral fits was that 
there is no spectral variability as the \ROSAT photon index was 
set equal to the \ASCA index. 
A third possibility is that the soft flux has indeed 
lower normalization as it could leak through a partial coverer 
or it could be scattered emission. 
This geometry probably applies to obscured AGN (e.g. Turner et al. 1997). 
According to this standard model, the soft X-rays are obscured from view 
by a thick torus while the hard X-rays 
can penetrate through the absorbing material. 
Then the soft X-rays  could be scattered emission 
on a pure electron medium. Then the scattered  power-law 
 should have the  same slope as the hard power-law. 
The best fit results for this model are given in table 2: 
We  see that we obtain a good $\chi^2$ compared to the 
single power-law model, where the 
\ASCA and \ROSAT normalizations are tied together.
Using the F-test we find that the change in the $\chi^2$ is 
statistically significant at over the 99.9 confidence level. 
The best-fit model favours a steep power-law, $\Gamma=2.1_{-0.17}^{+0.15}$, 
with a  large column density, $\rm N_H=35_{-13.2}^{+42.5}\times 10^{22}$ 
$\rm cm^{-2}$. The covering fraction is around 90 per cent, 
comparable to the values obtained in intermediate type Seyfert galaxies, 
e.g. Seyfert 1.9 (e.g. Turner et al. 1997). 
In Fig. \ref{cc2}  we present the confidence contours for the photon index 
and the column density for the model above while in Fig. \ref{xspec}
we give the  folded spectrum together with the data residuals
 from the fitted model (the data have been binned for clarity).

\section{DISCUSSION}
The broad $H_{\alpha}$ line detected shows that this 
object is an obscured AGN. Therefore this QSO
may be a bright example of a type-II QSO population which 
avoids detection in the optical surveys. Interestingly,  
this object  presents UV excess but it 
would not be classified as a QSO as it does not 
present broad lines in its optical spectrum. 
There is no $H_\beta$ line detected and this  
implies a reddening of at least $A_V > 3$.
Note however, that the reddening cannot be much higher than 
the value above; if   
 $A_V>100$ as the X-ray spectral fits suggest, 
 then the $\rm H_{\alpha}$  line would be 
obliterated as well.
This is consistent with the moderately red colour 
($B-K=5.4$) of this object; here we assumed that  
the relative AGN and host galaxy contributions 
are the same in the $B$ and the $K$ band. 
It is therefore a puzzle how the UV continuum 
is emerging unattenuated. One possibility 
is that a powerful star-forming galaxy hosts the QSO. 
Then, the QSO UV emission is  completely obscured  
due to large amounts of dust while we are observing the 
UV continuum of the starburst. Even in this case, the 
highly ionized C{\sc iv} ($\lambda1549$) and $\rm Ly_\alpha$ lines  
 cannot originate from star-forming processes.  
Another possibility is that the UV emission we observe 
is scattered light from the nucleus (Kinney et al. 1991). 
Then however, as the scattered fraction is usually of the order of 
a few percent, the unobscured 
total UV emission would be abnormally high.

Our  \ROSAT data suggested a hard spectrum of $\Gamma \sim1.56$. 
The \ASCA data alone confirm the \ROSAT findings. 
They show a flat spectrum, although it is 
difficult to disentangle whether this is due to  
a flat power-law or a large absorbing column.  
The joint \ROSAT-\ASCA spectrum again favours a hard spectrum.
Although a single power-law with high intrinsic 
absorption ($\Gamma=1.47$, $N_H\sim 10^{23}$ \cunits), 
cannot be strongly rejected,
the data prefer a \ROSAT power-law normalization 
below the \ASCA normalization. 
This could be attributed to variability between 
the \ASCA and the \ROSAT epochs. 
Indeed, there is some tentative ($2\sigma$) evidence for 
variability even between the two \ROSAT observations
 (see Almaini et al. 1998). 
If this is the case then a single flat power-law 
model with no intrinsic absorption would  fit the data. 
The best fit power-law spectrum is again hard 
($\Gamma =1.35^{+0.16}_{-0.14}$) very different from the 
spectrum of radio-quiet QSOs in \ASCA which have 
$\Gamma =1.93 \pm 0.06$ (Reeves et al. 1997).  
Alternatively, if we do not allow for variability 
between the \ASCA and the \ROSAT data 
we can fit the spectrum using a ``scattered'' power-law 
model i.e. adding a soft power-law 
 component with slope identical to that of the hard power-law.  
 The amount of the hard flux relative to the 
  scattered flux is given by the covering factor. 
 This model describes well the X-ray spectra of local 
obscured Seyfert galaxies (e.g. Seyfert 1.9 galaxies).   
We obtain  a best fit column density of 
$N_H\sim 30\times 10^{22}$ $\rm cm^{-2}$  while the 
spectral slope is $\Gamma=2.10^{+0.15}_{-0.17}$;  
  the covering factor is high ($f=0.9$).  
These values are consistent with those obtained for  
local obscured Seyfert galaxies (Turner et al. 1997). 
There may be a problem  however with this interpretation. 
The derived column density ($N_H=30\times 10^{22}$ $\rm cm^{-2}$ 
corresponds to $A_V\sim 180$ according to Bohlin et al. 1978) is much higher 
than the column inferred from the Balmer decrement ($A_V>3$); 
note that although the Balmer decrement gives 
only a lower limit to the visual extinction, $A_V$ 
cannot be much greater than the above value of 3 magnitudes, as then 
the $H_\alpha$ line would be obliterated.  
The same discrepancy is observed in some Seyfert 2 galaxies 
(e.g. Veilleux et al. 1997; Simpson 1998). 
Veilleux et al. suggest several possibilities 
to resolve this. The broad lines may originate 
at much larger distances from the nucleus where 
the X-ray emitting region is probably situated. 
Then the X-rays would view a larger column compared to that 
viewed by the broad-line region. Alternatively, as the X-rays are absorbed 
by heavy elements while the optical light is absorbed 
by dust, a large gas-to-dust ratio or a 
metallicity much higher than solar 
could alleviate the discrepancy.  
 For example, the former would be the case in the model 
of Granato, Danese \& Franceschini (1997) where the absorbing column 
is inside the dust sublimation radius.

A few other obscured AGN have been detected in the \ASCA surveys  
but at lower redshift ($z<1$) and only on the basis of 
high excitation narrow lines (Ohta et al. 1996; Akiyama et al. 1998;
Boyle et al. 1998). 
This population does not come as a great surprise. 
It was well-known that some narrow-line radio galaxies 
present broad emission lines in the IR and therefore they are 
hidden radio-loud QSOs (Economou et al. 1995; Hill et al. 1996).
These objects tend to have flat X-ray spectra 
probably due to an additional beamed component (e.g. Reeves et al. 1997).   
However, our object is not a powerful radio emitter.    
It is not detected in the NVSS survey (Condon et al. 1998),
thus having flux less than 2.5mJy at 1.4 GHz; 
this translates to a radio-optical spectral index 
(Zamorani et al. 1981) of $a_{ro}<0.4$. 
Although deeper radio observations are necessary in order to 
measure its radio power,  it is likely that our object 
represents  the first radio-quiet 
counterpart of these high redshift radio galaxies. 
The hard X-ray spectrum of this object,
similar to the spectrum of the XRB in the 2-10 keV band 
(e.g. Gendreau et al. 1995),  
suggests that such an obscured AGN population may 
make a significant contribution to the XRB.

\section{conclusions}
We have presented near-IR and hard X-ray (\ASCA) spectra of the obscured,
high-redshift, $z=2.35$, AGN RXJ13334+0001.
This object presents only narrow emission lines in its rest-frame UV 
spectrum. However,  a broad $H_\alpha$ line in the 
UKIRT spectrum  clearly classifies this object as a QSO. 
The absence of a broad $H_\beta$ line suggests a moderate amount of 
reddening (not much higher than $A_V\sim 3$ otherwise 
the $H_\alpha$ line would be obliterated as well) in agreement 
with the red colour ($B-K=5.4$) obtained with IRCAM3. 

The \ASCA data favour a flat or absorbed spectrum. 
The best-fit spectrum has $N_H\sim 10^{23}$ \cunits 
with the power-law  being rather 
unconstrained, $\Gamma = 1.9^{+0.62}_{-0.46}$. 
Combining the \ROSAT with the \ASCA data, keeping the 
\ASCA equal to the \ROSAT normalization, we obtain 
again a similar hard spectrum with $N_H\sim 10^{23}$ \cunits,
$\Gamma \sim 1.47^{+0.43}_{-0.20}$.
However, the above model is acceptable at only the 6 per cent 
confidence level. 
A (statistically significant) better fit is 
obtained when we use a scattered power-law model
(hard power-law plus scattered soft power-law). 
The best fit spectrum shows a high absorbing column 
$N_H\sim 3\times 10^{23}$ \cunits with $\Gamma =2.1^{+0.15}_{-0.17}$
and a covering fraction of 0.9. 
This is very similar to the spectrum of obscured 
Seyfert nuclei. 
However, the derived column density is far above 
the value obtained from the UKIRT data alone.
 This discrepancy is not uncommon in obscured Seyfert galaxies 
and it can be resolved on the basis of eg 
high gas metallicities, high gas-to-dust ratios,
or simply considering that 
the X-rays originate from the nucleus while the broad line region 
is located further away.    
 Alternatively, the discrepancy between the 
columns viewed in X-rays and in the IR can be 
alleviated if we allow for variability between the 
\ASCA and the \ROSAT epoch. 
Then the best fit X-ray spectrum 
is $\Gamma =1.35^{+0.10}_{-0.14}$ while there is no 
evidence for a large absorbing column 
 in better agreement with the UKIRT findings. 
 
Our object may be one of the first examples 
of a radio-quiet, type-II QSO.  
This population with properties similar to those 
of obscured Seyfert galaxies locally, remains undetected in 
optical UVX surveys. However, it could be readily detected in 
hard X-ray, the far-IR and mm part of the spectrum  
where obscuration effects are less important.  
Indeed, there are now several examples of 
obscured type-II AGN in the \ASCA surveys.  
Future surveys with the AXAF and XMM X-ray missions 
are expected to reveal large numbers  of these 
obscured AGN and to provide much better 
constraints on their X-ray spectrum and 
number density.

\section{Acknowledgements}
We thank the referee X. Barcons for his usefull suggestions. 
We thank S.J. Rawlings (Univ. of Oxford) for his assistance 
in the taking and reducing of the UKIRT CGS4 data. 
IG thanks the Institute for Astronomy at Edinburgh for their 
hospitality. We also thank A. Lawrence, J. Dunlop and 
C. Done for many useful discussions.
This research has made use of data obtained through the HEASARC
 online services, provided by the NASA Goddard Space flight Center. 
The optical data were obtained at the 
Anglo-Australian and the UKIRT telescope.

\section*{References}
Akiyama, M. et al., 1998, ApJ, 500, 173 \\
Almaini, O., Boyle, B.J., Griffiths, R.E., Shanks, T., Stewart, G.C., 
Georgantopoulos, I., 1995, MNRAS, 277, L31 \\
Almaini, O., Shanks T., Boyle B.J., Griffiths R.E., Roche N., Stewart G.C. 
\& Georgantopoulos I., 1996, MNRAS 282, 295 \\ 
Almaini, O., et al., 1998, MNRAS, submitted \\ 
Baker J.G. \& Menzel D.H., 1938, ApJ, 88, 52 \\
Barvainis R., 1993, ApJ, 412, 513 \\
Bevington P.R. and Robinson D.K., 1992, Data reduction and error analysis 
for the physical sciences, 2nd ed. \\
Bohlin, R.C., Savage, B.D., Drake, J.F., 1978, ApJ, 224, 132 \\
Boyle, B.J., Almaini, O., Georgantopoulos, I., Stewart, G.C., Shanks, T., 
Griffiths, R.E., Gunn, K., 1998, MNRAS, 297, L53 \\
Comastri, A., Setti, G., Zamorani, G., Hasinger, G., 1995, AA, 296,1 \\
Condon, J.J., Cotton, W.D., Greisen, E.W., Yin, Q.F., Perley, R.A.,
Taylor, G.B., Broderick, J.J., 1998, AJ, 115, 1693 \\ 
Economou, F., Lawrence, A., Ward, M.J., Blanco, P.R., 1995, MNRAS, 272, L5 \\
Gendreau, K., 1995, Ph.D. thesis, MIT \\
Georgantopoulos, I., Stewart, G.C., Shanks, T., Boyle, B.J., Griffiths, R.E.,
1996, MNRAS, 280, 276 \\ 
Georgantopoulos, I., Stewart, G.C., Blair, A.J., Shanks, T., Griffiths, R.E., 
Boyle, B.J., Almaini, O., Roche, N., 1997, MNRAS, 291, 203 \\
Granato, G.L., Danese, L., Franceschini, A., 1997, ApJ, 486, 147 \\
Halpern, J.P., Eracleous, M., Forster, K., 1998, ApJ, in press \\
Hasinger, G., Burg, R., Giacconi, R., Schmidt, M., Trumper, J., Zamorani, G., 
1998, AA, 329, 482 \\
Hill, G.J., Goodrich, R.W., De Poy, D.L., 1996, ApJ, 462, 163 \\
Iwasawa, K., Fabian, A.C., Brandt, W.N., Crawford, C.S., Almaini, O., 1997,
MNRAS, 291, L17 \\
Kinney, A.L., Antonucci, R.R.J., Ward, M.J., Wilson, A.S., 
Whittle, M., 1991, 377, 100 \\
Madau, P., Ghisellini, G., Fabian, A.C., 1994, MNRAS, 270, L17 \\
McHardy, I.M., et al., 1998, MNRAS, 295, 641 \\
Nandra, K., Pounds, K.A., 1994, MNRAS, 268, 405 \\
Ohta, K., Yamada, T., Nakanishi, K., Ogasaka, Y., Kii, T., Hayashida, K., 
1996, ApJ, 458, L57 \\
Osterbrock D.E., 1981, ApJ, 249, 462 \\
Reeves, J.N., Turner, M.J.L., Ohashi, T.,  Kii, T., 1997, MNRAS, 292, 468 \\
Shanks, T., Georgantopoulos, I., Stewart, G.C. Pounds, K.A., Boyle, B.J., 
Griffiths, R.E., 1991, Nature, 353, 315 \\ 
Shanks, T., Almaini, O., Boyle, B.J., Done, C., Georgantopoulos, I., 
Griffiths, R.E., Rawlings, S., Stewart, G.C., 1995, Spectrum, 7, 7 \\
Shanks, T., Almaini, O., Boyle, B.J., Della-Ceca, R., Done, C., 
 Georgantopoulos, I., Griffiths, R.E., Rawlings, S.J., Roche, N., 
 Stewart, G.C., 1996, In Roentgensstrahlung from the Universe, MPE Report 263, 
 Eds. Zimmermann, H.U., Trumper, J.E. and Yorke, H., p. 341  \\ 
Simpson, D., 1998, ApJ, in press  \\
Stark A. A. \etal, 1992, ApJS, 79, 77\\
Tanaka Y., Inoue H., Holt S.S., 1994, PASJ, 46, 37L \\
Tashiro O. et al., 1996, PASJ, 438, 157 \\
Truemper J., 1983, Adv. Sp. Res., 2, 241  \\
Turner, T.J., George, I.M., Nandra, K., Mushotzky, R.F., 1997, ApJ, 488, 164 \\
Veilleux S., Osterbrock D.E., 1987, ApJS, 63, 295 \\
Veilleux, S., Goodrich, R.W., Hill, G.J., 1997, ApJ, 477, 631 \\ 
Webster, R., Francis, P.J., Peterson, B.A., Drinkwater, M.J., Masci, F.J.,
1995, Nature, 375, 469 \\ 
Yaqoob T. \etal, 1997, The \asca ABC Guide, v2.0, NASA/GSFC  \\
Zamorani, G. et al., 1981, ApJ, 245, 357 \\

\end{document}